Challenges of Blockchain Applications in Digital Health: A Systematic Review

Andrew M. Nguyen MS, NREMT

Department of Public Health and Community Medicine, Tufts University School of Medicine, Boston, MA, USA

**1.1 Abstract**

Digital health, an emerging field integrating digital technologies into healthcare, is rapidly evolving and holds the potential to transform medical practices. Blockchain technology has garnered significant attention as a potential solution to various issues within digital health, including data security, automation, interoperability, and patient data ownership. However, despite the numerous advantages, blockchain faces several challenges and unknowns that must be addressed. This systematic literature review aims to explore the challenges of blockchain applications in digital health and provide best practices to overcome current and future roadblocks. Key issues identified include regulatory compliance, energy consumption, network effects, data standards, and the accessibility of the technology to stakeholders. To ensure the successful integration of blockchain within digital health, it is crucial to collaborate with healthcare stakeholders, pursue continued research and innovation, and engage in open discussions about the technology's limitations and potential.

**2.1 Introduction**

Digital health is the integration of digital technologies in healthcare (The Lancet Digital Health, 2021). Though still nascent, the emerging ecosystem is a multibillion dollar field that is formally recognized and studied by the American Medical Association, the US Food and Drug Administration, and the World Health Organization (Fatehi et al., 2020). Notable digital health technologies, such as telehealth, mobile health, and remote patient monitoring are becoming core elements of current and future medical practices (Ichikawa et al., 2017). These technologies have the potential to improve patient outcomes, support clinical decision making, and transform biomedical research (Shukla et al., 2022). In particular, digital health has become intertwined with ancillary healthcare domains, such as big data, artificial intelligence (AI), and internet of things (IoT) (Fatoum et al., 2021; Saravanan et al., 2017). Another emerging technology of interest within the field of digital health is blockchain.

Blockchain is a shared database of transactions originally created for financial applications (Kuo et al., 2019; Nakamoto, 2008). To append transactions to the database, or "ledger" (an aggregate record of all past transactions), "nodes", or computers, take turns adding data depending on a consensus algorithm, such as competing to solve cryptographic, computationally-intensive puzzles called hash functions (Elangovan et al., 2022). The node that solves the hash function first acquires the hash to the hash function (Azaria et al., 2016). That hash, along with the timestamp and metadata of the transaction, is appended to the ledger, creating an immutable "chain" of data "blocks" (Liang et al., 2017). All nodes then verify and validate that hash with their copy of the ledger, creating a distributed and decentralized database (Fatoum et al., 2021). Blockchains can also offer "smart contracts", which are digital code within the ledger that can autonomously execute pending pre-specified conditions (Lee et al., 2022).

Blockchains have the potential to solve various issues in digital health because they: 1. may be more secure and tamper-proof than current centralized medical record systems (Kuo et al., 2019); 2. can automate care processes such as insurance claims and remote patient monitoring without a third party mediator via smart contracts, which can in turn reduce costs and errors (Esmaeilzadeh, 2022); 3. provide data provenance and data auditing (Ichikawa et al., 2017; Azaria et al., 2016); 4. can serve as open, decentralized, and interoperable data lakes that facilitate large-scale data transfers needed to power big data, AI, and IoT applications (Saravanan et al., 2017; Shukla et al., 2022; Fatoum et al., 2021); 5. can allow patients the ability to proactively consent and authorize third parties to access, share, and use their data (Azaria et al., 2016); 6. may promote patient self-sovereignty, self-advocacy, and self-ownership of data (Liang et al., 2017; Kostick-Quenet et al., 2022); and 7. may secure personal health information to patients as a civil and human right (Costa et al., 2022).

To further explore the potential of blockchain, many research organizations, healthcare systems, and even countries have studied its application in digital health. In 2017, Saravanan et al. used smart contracts with wearables to create an end-to-end blockchain-based protocol to remotely and securely monitor patients with diabetes. Similarly, Ichikawa et al. developed and evaluated an mHealth app for cognitive behavioral therapy treatments for insomnia using blockchain (2017). Shukla et al. expanded on the interoperability between IoT medical devices and EHR systems through a blockchain Application Programming Interface (API) (2022). Estonia adopted a blockchain-based healthcare platform for patient health records in 2016 (Yun et al., 2020).

Though many studies have emphasized the benefits of blockchain applications in healthcare, few have comprehensively analyzed the challenges that current projects face, especially in the context of digital health. Evangelatos et al. offered a review of prospects and challenges, but did not pursue an extensive literature review (2020). To our knowledge, there is no thorough review on the challenges of blockchain in digital health in the literature. Therefore, there is a knowledge gap in our understanding of how blockchain can best fit the needs of digital health solutions. To bridge this gap, we propose a systematic literature review of challenges of blockchain applications in digital health and offer best practices to overcome current and future roadblocks.

## 3.1 Methods

We surveyed bibliographic databases from PubMed spanning from 2016 to 2022 and queried searches for "blockchain" or "blockchain digital health" or "blockchain digital" or "blockchain health" or "blockchain records" or "blockchain health records." We surveyed 87 articles and selected 12 relevant studies mentioning blockchain applications in digital health along with any technical, social, and cultural challenges. After a thorough review of selected articles, challenges were identified and organized into 11 themes (Table 1).

Results

Many articles (4/12, 33%) described the desire for further research due to the nascent blockchain industry. In particular, Azaria et al. cited the need for political and legal regulation, especially in the context of healthcare (2016). Clarity is also needed on its implementation and integration with legacy systems (Ichikawa et al., 2017). Many studies also described a lack of clinical studies demonstrating how blockchain can affect patient outcomes, satisfaction, or costs (Fatoum et al., 2021). Likewise, the availability of large-scale datasets or blockchain health systems to elucidate these metrics also remain absent (Yun et al., 2020).

The intrinsic nature of some blockchain consensus mechanisms, such as proof-of-work, which involves the process of nodes competing to solve intensive cryptographic puzzles, may demand higher energy costs to maintain than traditional centralized databases (Fatoum et al., 2021). The massive computing power required to scale blockchain networks would likely exclude stakeholders participating with low-power IoT devices (Anagnostakis, et al., 2021). To even store such data on-chain, enormous storage solutions would also be needed (Evnagelatos, 2020).

Corollary to the computational and energy-intensive nature of some blockchains is the concern of scalability and throughput. The most cited challenge (6/12, 50%) was the ability to effectively transact between billions of IoT devices (Anagnostakis, et al., 2021). As transaction volume increases, so too does the power needed to verify transactions (Kostick-Quenet et al., 2022). In fact, Fatoum et al. reported that the Hyperledger blockchain could perform at most 800 transactions per second, far below the needs of even a small health system (2021).

**4.1 Discussion**

These unknowns, along with the possibility to potentially disrupt current business and care models, pose a risk to further blockchain endeavors. Esmaeilzadeh et al., suggest that deliberately working alongside regulators and ensuring compliance with current standards, such as HIPAA, may ease these concerns as adoption and research continues (2022).

Because of these costs, most blockchains do not actually hold patient data, but rather the metadata to it. Keeping patient data off-chain effectively alters the retrieval and sharing process, but leaves that off-chain data susceptible to the same weaknesses of traditional, centralized systems (Yun et al., 2020). At the same time, the granularity and specificity of healthcare data as a "digital fingerprint" may render the pseudonymity features of blockchain useless, as the open nature of on-chain patient data can be used to identify them if not encrypted (Kostick-Quenet et al., 2022). Alternatively, energy-efficient consensus mechanisms, such as proof-of-stake, may resolve issues of computing power, while federated blockchain models may offer both the security of blockchain and storage capacity of traditional health systems (Costa et al., 2022;

Esmaeilzadeh, 2022). Research in data encryption techniques should also continue to be explored (Lee et al., 2022).

To address issues on network effects, developers should involve stakeholders in blockchain implementation and educate them on the benefits and risks (Esmaeilzadeh, 2022).

Beyond technical limitations, the ubiquitous challenge of data standards remains. Without proper planning, blockchains may further perpetuate data silos without general standards or protocols for sharing data between chains or health systems (Yun et al., 2020). Legal and ethical issues emerge from smart contracts that may autonomously execute the rights to share, sell, or view data (Kostick-Quenet et al., 2022). If patient data and personal information is held on-chain and is unencrypted, they may also be accessed by third parties who are not directly involved in patient care (Ichikawa et al., 2017). Therefore, blockchain development must be considered in parallel with current standardized medical data formats, such as Health Level 7 Fast Healthcare Interoperability Resources (Lee et al., 2022).

Furthermore, the introduction of new, complex technologies excludes stakeholders who may not have the knowledge or resources to participate. Kostick-Quenet et al. argue that patients may yield their rights to trusted, centralized intermediaries to maintain their data, thus defeating the purpose of blockchain applications (2022). Involving healthcare stakeholders in blockchain development end-to-end and educating them on its function can prompt discussions in resolving legal or ethical issues (Esmaeilzadeh, 2022). Until blockchains overcome these challenges, the technology may not be suitable for the velocity, volume, and veracity of data applications in digital health (Azaria et al., 2016).

**5.1 Conclusion**

Blockchain is a promising solution to bridge data between digital health solutions (big data, AI, IoT) and traditional health systems. It has the potential to resolve issues on health data security, automation, provenance, interoperability, patient consent, ownership, and civil and human rights. However, it is still an emerging technology with a plethora of technical, social, and cultural unknowns and challenges. Its ability to improve care and offer value is dependent on collaboration with existing healthcare stakeholders, continued research and innovation, and further discussion and inquiry.

**Appendix**

| | |
|---|---|
| Azaria et al., 2016, Ichikawa et al., 2017, Esmaeilzadeh, 2022, Yun et al., 2020 | Nascent industry (4/12, 33%) |
| Fatoum et al., 2021, Yun et al., 2020 | Lack of studies showing clinical outcomes (2/12, 16%) |
| Evenelatos et al., 2020; Kostick-Quenet et al., 2022; Fatoum et al., 2021; Anagnostakis et al., 2021 | Cost and energy consumption (4/12, 33%) |
| Fatoum et al., 2021, Azaria et al., 2016, Kostick-Quenet et al., 2022, Lee et al., 2022; Evenelatos et al., 2020; Anagnostakis et al., 2021 | Scalability and throughput (6/12, 50%) |
| Kostick-Quenet et al., 2022, Yun et al., 2020; Evenelatos et al., 2020 | Only metadata is "on-chain" (3/12, 25%) |
| Ichikawa et al., 2017 | Consensus attack (1/12, 8%) |
| Esmaeilzadeh, 2022 | Reliance on network effects (1/12, 8%) |
| Kostick-Quenet et al., 2022, Yun et al., 2020, Saravanan et al., 2017 | Pseudonymity, privacy (3/12, 25%) |
| Lee et al., 2022, Kostick-Quenet et al., 2022, Yun et al., 2020 | Data standards (3/12, 25%) |
| Azaria et al., 2016, Esmaeilzadeh, 2022, Kostick-Quenet et al., 2022; Ichikawa et al., 2017 | Legal and ethical implications (4/12, 33%) |
| Kostick-Quenet et al., 2022 | Socioeconomic exclusion (1/12, 8%) |

Table 1. Themes identified in 12 research articles.